# Breast Cancer Detection Using Deep Learning Technique Based On Ultrasound Image

Abdulqader Mohammed[1]    Mohammed Abdel Razek[1]    Mohamed El-dosuky[2,3]    Ahmed Sobhi[1]

1 Department of Mathematics and Computer Science, Faculty of Science, AL-Azhar University, Nasr City, Cairo11884, Egypt.
2 Department of Computer Science, Faculty of Computers and Information, Mansoura University, Egypt.
3 Department of Computer Science, Arab East Colleges, Saudi Arabia.

**Abstract:**

Breast cancer ranks as the most prevalent form of cancer diagnosed in women, and diagnosis faces several challenges, a change in the size, shape and appearance of breasts, dense breast tissue, lumps or thickening in the breast especially if in only one breast, lumps and nodules in the breast. The major challenge that faces deep learning diagnosis of breast cancer was its shape, size and position non-uniformity especially malignant cancer. This work proposed a deep learning system that increased the accuracy of classification of breast cancer types from ultrasound images. It reaches 99.29% accuracy, exceeding other previous work. First, image processing was applied to in enhance the quality of input images. Second, the image segmentation was performed using U-Net architecture. Third, many features are extracted using Mobilenet. Finally, the accuracy of proposed system was evaluated.

## 1. Introduction

Breast cancer is a common type of cancer that forms in the breast cells and is more common in women than in men [1]. Age, gender, family history, and genetic factors contribute to the risk. mutations, hormonal factors, personal history, and lifestyle factors [2]. Symptoms include a lump or mass in the breast, breast pain, swelling, skin irritation, nipple retraction, and redness [3]. Diagnosis and treatment involve breast examination, imaging tests, and biopsy [4]. Available treatment options encompass surgery, radiation therapy, chemotherapy, hormone therapy, and targeted therapy [5]. Prevention and awareness of breast cancer include regular self-exams, clinical breast exams, mammography screening, and a healthy lifestyle [6].

Breast cancer diagnosis faces several challenges, including false positives and false negatives, dense breast tissue, subjective interpretation, limited access to screening, over diagnosis and overtreatment, lack of standardized guidelines, and invasive diagnostic procedures [7].

Deep learning, which is a subset of machine learning, has shown promise in breast cancer diagnosis and treatment [8]. These algorithms can analyze mammograms, ultrasound images, and MRI scans to detect and classify breast lesions, improve image interpretation accuracy, and predict individual risk of developing breast cancer.

Contributions, this paper proposed a deep learning model that increased the accuracy of classification of breast cancer from ultrasound images. It reaches 99.29% accuracy, exceeding other previous work. First, image processing was applied to improve the clarity and resolution of input images. Second, the U-Net architecture was utilized for the purpose of segmentation. Third, many features are extracted using Mobilenet. Finally, the classification using VGG 16 accuracy of proposed system was evaluated.

The subsequent sections of this paper are organized as follows. Section 2 provides an overviews of the previous work in breast cancer diagnosis. Section 3 provides the proposed methodology. Section 4 shows the implementation. Section 5 shows results. Finally, section 6 conclusion of the paper before listing some future directions.

## 2. Previous work

Table 1 shows the previous work on breast cancer diagnosis based on ultrasound images along with accuracy. In reference [9], proposed The combination of deep learning technology with ultrasound imaging diagnosis was employed. The tumor regions were segmented from the breast ultrasound (BUS) images using a supervised block-based region segmentation algorithm. The best diagnostic outcome was achieved by establishing a combination feature model based on the depth feature of ultrasonic imaging and strain elastography. In reference [10], proposed To tackle the issue of noisy labels during the training of breast tumor classification models, a successful technique known as the noise filter network (NF-Net) was introduced. In reference [11], proposed a CAD system was developed for tumor diagnosis. This system utilized an image fusion method that combined various image content representations and employed ensemble techniques with different convolutional neural network (CNN) architectures on ultrasound images. In reference [12], proposed A new BIRADS-SSDL network was proposed, which incorporated clinically-approved breast lesion characteristics (BIRADS features) into task-oriented semi-supervised deep learning (SSDL). This integration aimed to achieve precise diagnosis of ultrasound images, particularly when working with limited training data. In reference [13] proposed a CAD system was developed for tumor diagnosis. This system employed an image fusion method that combined various image content representations and utilized ensemble techniques with different convolutional neural network (CNN) architectures. It was observed that different image content representations had an impact on the prediction performance, with increased image information leading to improved prediction accuracy. Additionally, the inclusion of tumor shape features enhanced the diagnostic effectiveness of the system. In reference [14] proposed the breast ultrasound (BUS) images underwent resizing and were then enhanced using the contrast limited adaptive histogram equalization method. The pre-processed image was encoded using the variant enhanced block. Ultimately, the segmentation mask was generated through concatenated convolutions. In reference [15] A semi-supervised GAN model was created to enhance breast ultrasound images. The generated images were then employed for breast mass classification using a convolutional neural network (CNN). The performance of the model was assessed using a 5-fold cross-validation approach. In reference [16] The proposed method applies fuzzy enhancement and bilateral filtering algorithms for the enhancement of original images, obtain decomposition images representing breast tumor clinical characteristics, fuse them through RGB channels, choose the optimal deep learning feature model, and train a network for classification that utilizes adaptive spatial feature fusion technology. In reference [17] proposed a deep learning-based method for classifying breast masses in ultrasound images. The approach incorporates deep representation scaling (DRS) layers between pre-trained CNN blocks. By reducing the number of trainable parameters, this technique outperforms conventional transfer learning methods and achieves improved performance. In reference [18] proposed introduces transfer learning methods for the classification and detection of breast images in ultrasound. The focus is on transfer learning approaches, pre-processing techniques, pre-training models, and convolutional neural network (CNN) models. In reference [19] proposed a novel convolutional neural network with a coarse-to-fine feature fusion approach is suggested for breast image segmentation. The network comprises an encoder path, decoder path, and core fusion stream path, which collectively produce comprehensive feature representations for precise segmentation of breast lesions. Additionally, the network integrates super-pixel images and a weighted-balanced loss function to handle variations in lesion region sizes. In reference [20] proposed a BUViTNet, a method for breast ultrasound detection using vision transformers (ViTs) instead of convolutional neural networks (CNNs). The approach leverages datasets containing images from both ImageNet and cancer cells to classify breast ultrasound images. The performance of the algorithm surpassed that of ViT trained from scratch, ViT-based conventional transfer learning, and transfer learning based on CNN. In reference [21] the work presents a standard for breast ultrasound image segmentation evaluation, proposes standardized procedures for accurate annotations, and introduces a losses-based approach to assess the impact of user interactions on the sensitivity of semi-automatic segmentation. In reference [22] The study proposes a BI-RADS a classifier model for categorizing US breast lesions using a novel multi-class US image, utilizing bilinear interpolation and neighborhood component analysis to generate informative features for automated classification. In reference [23] The proposal introduces an enhanced ViT architecture that incorporates a shared MLP head to the output of each patch token. This modification ensures balanced feature learning between class and patch tokens. Additionally, the model utilizes the output of the class token to distinguish between malignant and benign images. Furthermore, the output of each patch token is employed to determine if the patch overlaps with the tumor area.

Table 1: Previous work.

| References | Technique | Year | Accuracy |
|---|---|---|---|
| [9] | Automatic identification using supervised block-based region segmentation and feature combination migration as the foundation | 2019 | 92.95 |
| [10] | learn from noisy ultrasound images | 2020 | 73.0 |
| [11] | Learning from a combination of convolutional neural networks | 2020 | 94.62 |
| [12] | Semi-supervised deep learning focused on BIRADS features | 2020 | 94.23 |
| [13] | Computer-aided diagnosis using ensemble learning from CNN | 2020 | 94.62 |
| [14] | Using variant-enhanced deep learning for ultrasound image segmentation. | 2021 | 89.73 |
| [15] | Semi-supervised Generative Adversarial Networks | 2021 | 90.41 |
| [16] | image decomposition and fusion | 2021 | 95.48 |
| [17] | transfer learning with deep representations scaling | 2021 | 91.5 |
| [18] | Transfer Learning in Breast Cancer Diagnoses | 2021 | 97 |
| [19] | Coarse-to-fine fusion convolutional neural network for breast image segmentation | 2021 | 97.17 |
| [20] | Breast ultrasound detection via vision transformers | 2022 | 95 |
| [21] | Benchmark Breast Ultrasound Image Segmentation | 2022 | 90 |
| [22] | Automated lesion BI-RADS classification using the pyramid triple deep feature generator technique | 2022 | 88.67 |
| [23] | Malignant Breast Ultrasound Images Using ViT-Patch | 2023 | 89.8 |

## 3. Proposed methodology:

This system utilizes four models: preprocessing, a segmentation model, a feature evaluation stage to extract various features from the cancer mask (contour), and cancer classification model.

In the first stage of the proposed system, its pre-processing the images to get rid of noise and improve them by applying the appropriate filter to them, then re-sizing and standardizing the size of all the images to be one size and format to deal with them accurately.

In the second phase, were segmented the images using the U-Net method, which is one of the deep learning methods and techniques and is effective in dealing with medical images, especially tumors. Image segmentation is an important stage, as we obtain from it the parts that contain the tumor in order to facilitate dealing with them and isolate the tumor area to extract characteristics from it.

The segmentation model uses the U-Net architecture and was trained to achieve 98.68% accuracy for training phase, and 99.29 accuracy for testing phase. After training, the actual and predicted masks were visualized for comparison.

In the third stage, were features extracted from the images using the Mobilenet method, which is a deep learning method that is accurate and efficient and reduces the number of parameters compared to other networks, resulting in deep and lightweight neural networks. These features were relied upon to determine the type of tumor whether it is malignant, benign or not cancer (normal). Many of these characteristics were extracted to identify the tumor accurately and efficiently, and they were taken from the mask that was made on the image, as it works to isolate the tumor area in order to take characteristics from it and know the tumor's sleep through those characteristics.

In the last stage came the classification phase, where one of the deep learning techniques, namely Visual Geometry Group VGG16, was used to know the type of tumor and identify it based on the features that were extracted from the mask in the third stage, where after extracting the features from the mask, they are entered into the deep network to train it with the network's layers and stages, and then the Identify and determined the type of tumor.

The classifier model was used to classify types of breast cancer, and the resulting classifications were visualized, using a Kaggle notebook instead of Google Colab because it has better GPUs.

Figure 1 shows the proposed architecture.

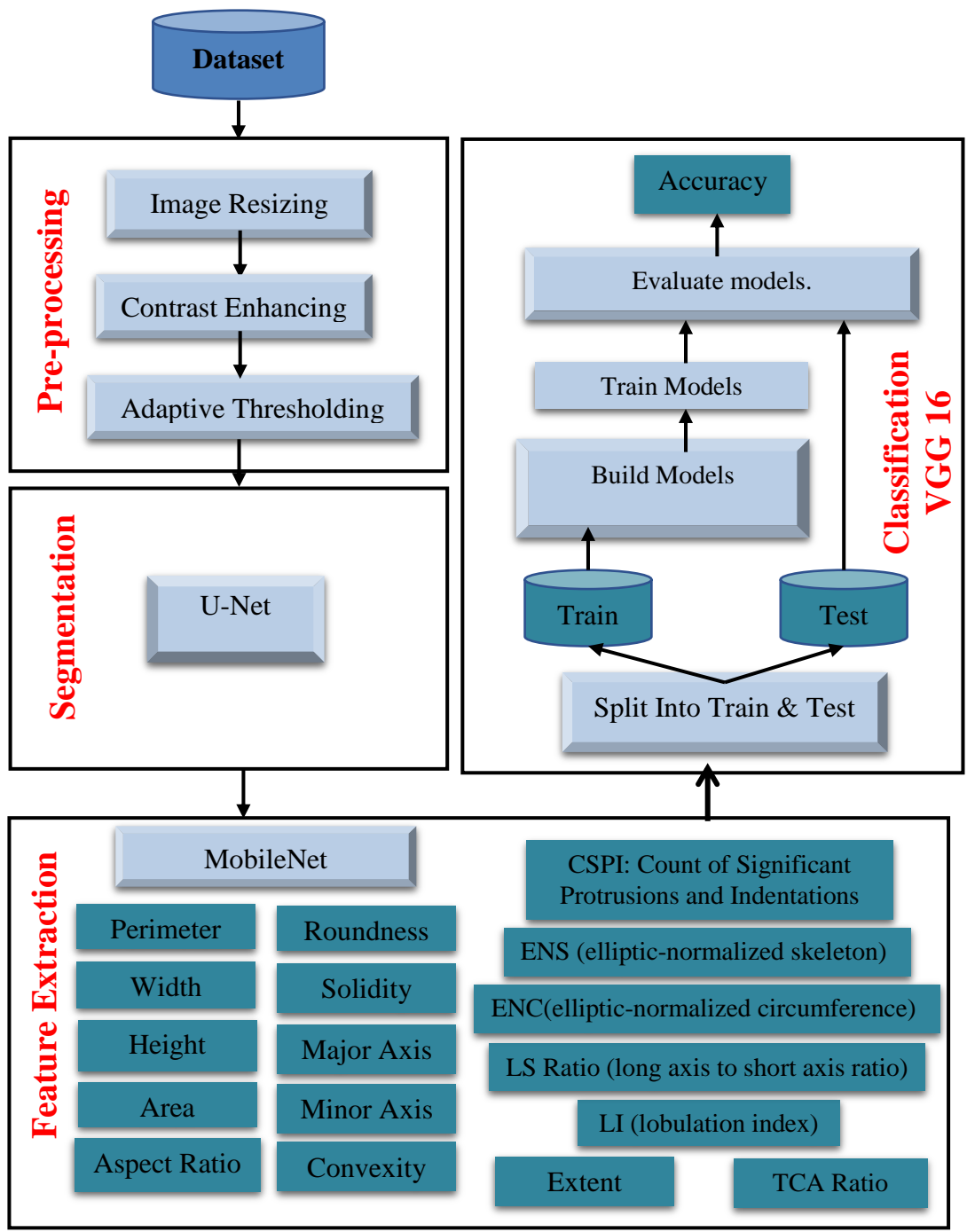

Figure 1: Proposed architecture.

Got Experience with image analysis and processing techniques, as well as deep learning algorithms for classification and feature extraction. The project involved the following tasks:

**3.1 Dataset:**

In 2018, data was collected at baseline, which includes breast ultrasound images of women aged 25 to 75. The dataset, obtained from Baheya Hospital for Early Detection & Treatment of Women's Cancer in Cairo, Egypt, is divided into three classes: normal, benign, and malignant images. The dataset comprises 780 images without a mask and 1583 images with a mask, with an average image size of 500×500 pixels. The images are in PNG format. Figure 2 shows Samples of breast images dataset.

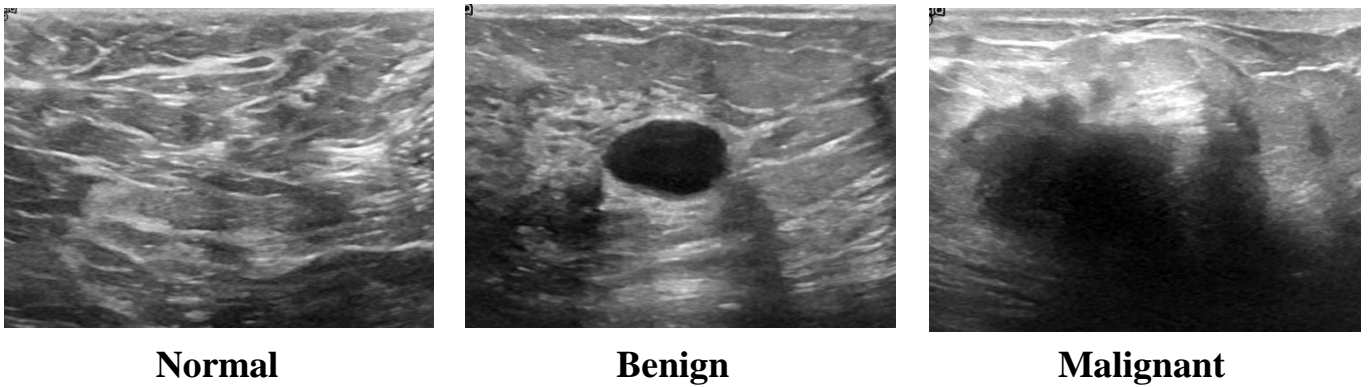

**Normal**         **Benign**         **Malignant**

Figure 2: Samples of Ultrasound breast images.

## 3.2 Pre-processing:

Data pre-processing and cleaning: Collecting and pre-processing the dataset of breast cancer images to prepare it for training and testing. The work first converts the colour format of the x-ray image from BGR to RGB and applies thresholding to create a binary image.

Detecting cancer and Evaluating the properties including the following steps:

1. File path for the image.

2. Load the image.

3. Convert the colour format from BGR to RGB.

4. Apply thresholding to create a binary image.

5. Detect contours in the binary image.

6. Display the image.

## 3.3 Segmentation:

Image segmentation involves dividing an image into multiple segments in digital image processing and computer vision [24]. The goal of this process is to transform image representations into meaningful objects to facilitate analysis. In the context of breast imaging and digital mammography, segmentation is commonly used for various purposes, including detecting breast contours [25] and identifying the pectoral muscle in the mediolateral oblique (MLO) view [26].

This work provided includes two models for a breast cancer x-ray images dataset: a segmentation model and a classification model.

U-net is a group of fully convolutional networks designed for segmenting objects in digital images. Its structure includes a downsampling path followed by an upsampling path, allowing [27]. It to utilize high-resolution lesion information from the shallow layer and provide missing detailed spatial information during the upsampling process, resulting in improved segmentation outcomes [28].

The U-Net architecture is commonly utilized in deep learning and was specifically created to tackle the issue of scarce annotated data in the medical domain. Its design allows for efficient utilization of smaller datasets without compromising on speed and precision [29].

The U-Net model, utilized for biomedical image segmentation, comprises three components: the encoder, the bottleneck module, and the decoder, and is classified as a Fully Convolutional Network (FCN). The U-Net, which is extensively employed, fulfills the needs of medical image segmentation due to its U-shaped architecture, incorporation of contextual information, rapid training speed, and efficient utilization of limited data [30].

The U-Net architecture stands out due to its distinctive structure, which includes both a contracting path and an expansive path. The contracting path is made up of encoder layers that capture contextual information and decrease the spatial resolution of the input. Meanwhile, the expansive path consists of decoder layers that decode the encoded data and utilize information from the contracting path through skip connections to produce a segmentation map [31].

U-Net uses a contracting path to identify features in input images, with encoder layers convolutional operations to increase depth and spatial resolution. The expansive path decodes encoded data while retaining input spatial resolution, with decoder layers performing convolutional operations and upsample feature maps. Skip connections retain spatial information for better feature location [32].

The segmentation model is a U-Net model, which is a kind of convolutional neural network designed for image segmentation tasks. The model is trained on a dataset of x-ray images and their corresponding masks of tumor contours.

The images are pre-processed and passed through the network, which outputs a predicted mask for each image. The model is then trained using the Adam optimizer and binary cross-entropy loss function. The model's performance is visualized using a line chart. Figure 3 shows U-Net architecture.

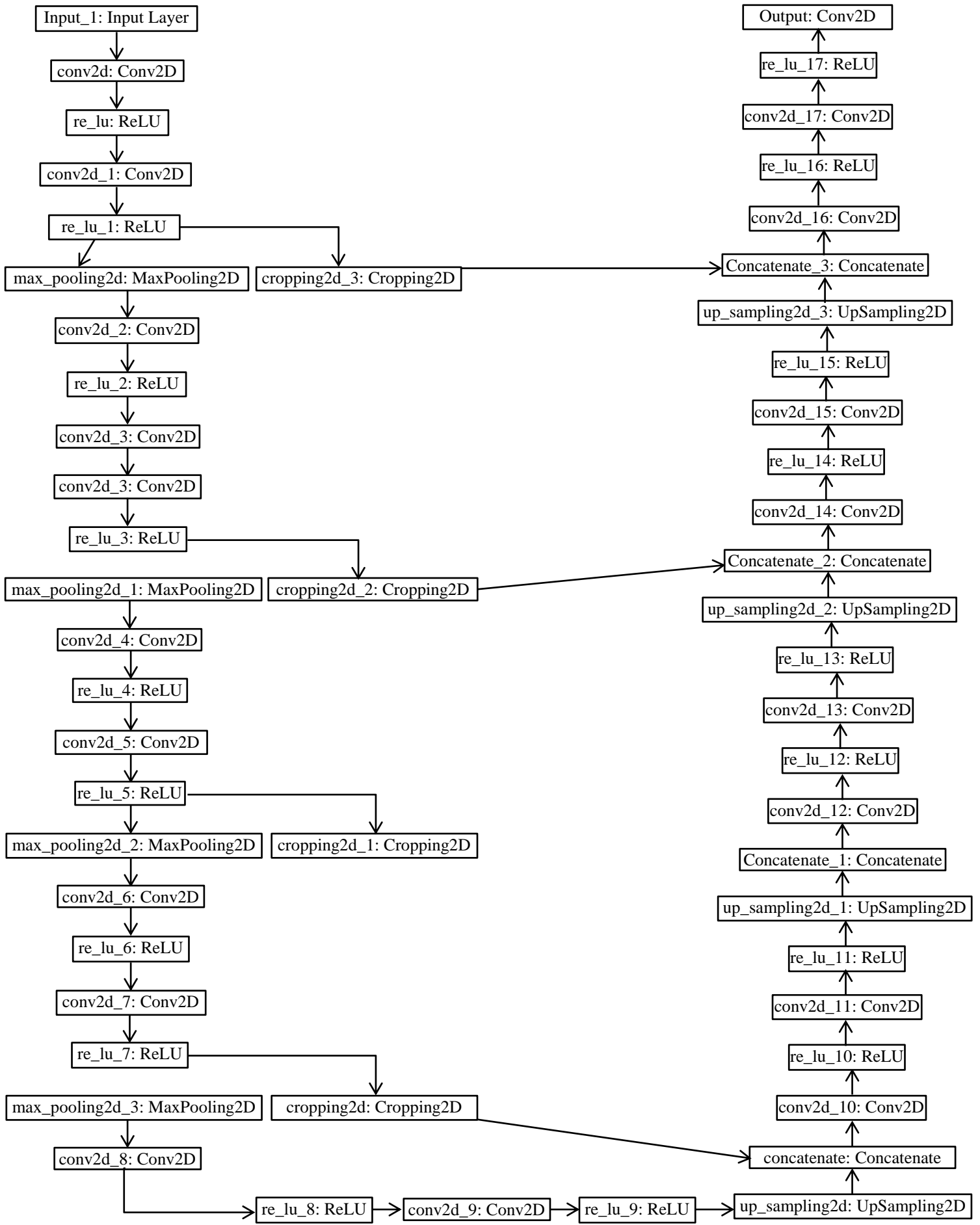

Figure 3: U-Net architecture

## 3.4 Feature extraction:

Feature extraction involves reducing the dimensionality of raw data to more manageable groups for processing. Large data sets often contain numerous variables that demand substantial computing resources. Feature extraction refers to methods that choose and/or merge variables into features, effectively decreasing the volume of data that needs to be processed, while still accurately and comprehensively describing the original data set [33].

In mammogram classification, feature extraction plays a crucial role in determining the classification outcome. The majority of systems utilize feature extraction to identify and categorize abnormalities as either benign or malignant based on texture, statistical properties, spatial domain, fractal domain, and wavelet bases [34].

Implementing feature extraction techniques such as MobileNet, inception V3 and other technique to extract relevant features from the breast cancer images.

The work also includes feature extraction for tumor contours. It then finds the contours in the binary image and draws them on the x-ray image.

MobileNet is a straightforward yet effective and relatively low-computational-requirement convolutional neural network designed for mobile vision tasks. It is extensively employed in various practical applications, such as object detection, fine-grained classifications, face attributes, and localization [35].

MobileNets rely heavily on depthwise separable convolutions, and used to minimize computation in the early layers. Flattened networks employ fully factorized convolutions to construct a network and demonstrate the potential of highly factorized networks [36].

The MobileNet model utilizes depthwise separable convolutions, which are a type of factorized convolutions that decompose a standard convolution into a depthwise convolution and a $1\times1$ convolution known as a pointwise convolution. In MobileNets, the depthwise convolution applies a single filter to each input channel, while the pointwise convolution combines the outputs of the depthwise convolution using a $1\times1$ convolution. Unlike a standard convolution, which filters and combines inputs into new outputs in a single step, the depthwise separable convolution divides this process into two separate layers for filtering and combining. This factorization significantly reduces computation and model size [37].

The area, perimeter, height, width, CSPI, aspect ratio, LI, ENS, ENC, LS ratio, convexity, extent, and TCA ratio, features of the contour are calculated and displayed and also fits an ellipse around the minimum area rectangle enclosing the largest contour and calculates the major and minor axis of the ellipse, roundness, and solidity [38][39]. The features are:

1. **Perimeter:** The Perimeter function measures the length of the tumor's boundary, which tends to be irregular in malignant tumors. A higher perimeter value is indicative of a higher probability of malignancy.

2. **Height:** Bounding Rectangle Height (BRH), the height of the smallest rectangular area that contains the region of interest (ROI).

3. **Width:** Bounding Rectangle Width (BRW), the width of the smallest rectangle that surrounds the ROI.

4. **Area:** The Area feature represents the size of a breast tumor, with malignant tumors often exhibiting a larger area in comparison to benign tumors.

5. **CSPI:** Count of Significant Protrusions and Indentations. The CSPI feature can be used to measure the extent of irregularity in the boundary.

   If $p_i$ is a point on the contour, the k-curve angle of $p_i$, denoted as $\theta_i$, can be calculated using $p_i$, $p_{i+k}$, and $p_{i-k}$, where k can be any defined number. The point $p_i$ is considered a smooth point if $\theta_i$ is $\leq 40°$, and is classified as a convex or concave point if $\theta_i$ is $> 40°$. If there are no concave points between any two convex points, the convex point with the smallest k-curve angle would be removed. Similarly, if there are no convex points between any two concave points, the concave point with the smallest k-curve angle would be removed.

   Figure 4 provides a demonstration of convex and concave points present in a tumor contour, followed by the definition of CSPI.:

$$\text{CSPI} = 2 \times n \qquad (1)$$

   Where n represents the count of concave points. Malignant lesions typically exhibit a jagged boundary, resulting in a higher CSPI for malignant tumors. Figure 4 shows the concave and convex points.

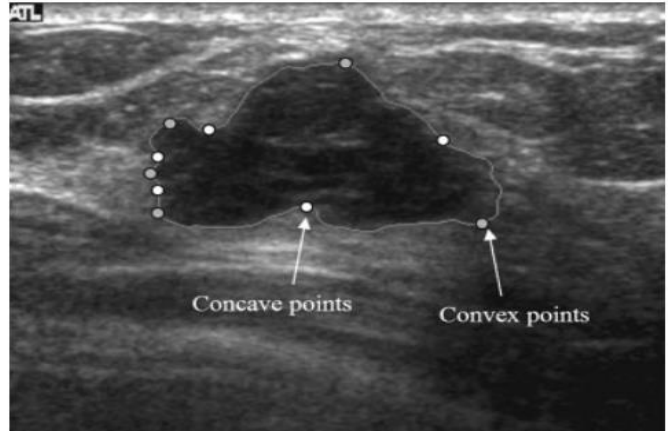

Figure 4: Illustration of convex points (gray) and concave points (white) within the contour of a malignant breast tumor.

6. **LI (lobulation index):** The lobe region bounded by a lesion contour and a line joined by any two neighboring concave points may be produced, in accordance with the definition for a concave point from the CSPI.

    If the sizes of the biggest and smallest lobe areas are represented by $A_{max}$ and $A_{min}$, and the average size of all lobe regions is represented by Average, then LI may be defined as follows:

    $$LI = \frac{(A_{max} - A_{min})}{A_{average}} \qquad (2)$$

    A malignant tumor often has a greater LI than a benign tumor.

7. **ENS (Elliptic-normalized skeleton):** A tumor region's skeleton expresses a set S, and ENS is the total of the points in S. The skeleton is also intricate in tumors with twisted boundaries. displays a sample of a cancerous tumor's skeleton. A malignant lesion always produces a significant ENS and has a twisted border. The internal lines in Figure 5 depict the skeleton of a malignant malignancy.

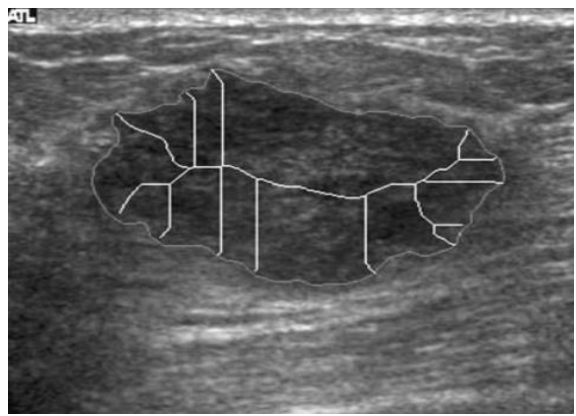

Figure 5: An illustration of a malignant breast tumor's skeleton (interior lines)

8. **Aspect Ratio:** This is the length ratio between the width and depth of a tumor. A tumor has a higher chance of being malignant if its depth is bigger than its breadth and its aspect ratio is more than 1.

$$\text{From factor} = \frac{4\pi \times \text{Area}}{\text{Perimeter}^2} \quad (3)$$

The tumor is almost spherical when the Form Factor is approaching 1.

9. **Roundness** $= \frac{(4 \times \text{Area})}{(\pi \times \text{Max Diameter}^2)} \quad (4)$

Where Max Diameter is the main axis length from the tumor's corresponding ellipse.

10. **Solidity** $= \frac{\text{Area}}{(\text{Convex\_Area})} \quad (5)$

Where solidity is near to 0, indicating that the tumor is malignant, and Convex size is the size of the convex hull of a tumor.

11. **Major Axis (MaA):** the principal axis of the ellipse that fits the ROI the best.

12. **Minor Axis (MiA):** the secondary axis of the ellipse that best fits the ROI.

13. **ENC (elliptic-normalized circumference):** The second-order moment may be used to determine each tumor's angle of inclination with regard to the x-y coordinate plane. It is thus possible to create an identical ellipse with the same area, center, and inclination angle for each tumor. The ENC may be defined as follows if Equivalent Ellipse Perimeter is the equivalent ellipse's perimeter:

$$\text{ENC} = \frac{(\text{Equivalent Ellipse Perimeter})}{(\text{Perimeter})} \quad (6)$$

A smooth tumor border indicates a high probability of benignity when the ENC value of a suspected breast tumor is around 1.

14. **LS Ratio:** The length ratio of the main long axis and minor short axis of the equivalent ellipse described in the ENC feature is known as the long axis to short axis ratio.

15. **Convexity** $= \frac{\text{Convex Perimeter}}{\text{Perimeter}} \quad (7)$

Where the convex hull of a tumor's perimeter is known as its convex perimeter.

16. **Extent** $= \frac{\text{Area}}{\text{Bounding Rectangle}} \quad (8)$

Where the smallest rectangle that contains the tumor is called the Bounding Rectangle.

17. Tumor area to convex area ratio, or TCA ratio, is expressed as follows:

$$\text{TCA Ratio} = \frac{Area}{Convex\ Area} \quad (9)$$

### 3.4 Classification Model:

The classification model uses utilizing transfer learning with a VGG16 model that has been pre-trained model. The pre-trained layers are frozen, and a new classification model is added on top. Dividing the data into an 80% Training set and a 20% Validation set.

### 3.4.1 Training and testing deep learning models:

Developing and testing machine learning models such as Visual Geometry Group VGG, and convolutional neural networks (CNNs) to accurately classify breast cancer images.

The VGG16 model is a 16-layer convolutional neural network (CNN) architecture known for its exceptional performance in computer vision tasks. The first convolutional layer uses a kernel size of 11, while the second layer uses a kernel size of 5. Rather than relying on a multitude of hyper-parameters, VGG16 employs 3x3 filters and stride 1 in its convolution layers, with consistent use of same padding and 2x2 stride 2 maxpool layers throughout the architecture. The network concludes with two fully connected layers and a softmax for output. The "16" in VGG16 signifies the 16 layers with weights. This network is quite substantial, containing approximately 138 million parameters [32].

The new model consists of a flatten layer, a dense layer with 256 units, and a concluding dense layer featuring a sigmoid activation function. It has been trained on a dataset containing x-ray images classified as either benign or malignant. The model's performance is assessed and presented using a confusion matrix.

After training, the actual and predicted masks were visualized for comparison. The classifier model was used to classify types of breast cancer, and the resulting classifications were visualized.

### 4. Implementation

The authors used Kaggle to edit and run the code, which was built in Python version 3.11. Kaggle, founded in 2010 and acquired by Google in 2017, is an online platform for data science and machine learning competitions. It offers a community-driven environment for data scientists and enthusiasts to explore, analyze, and solve real-world data problems. Users can share code, notebooks, and insights related to data science projects on its cloud-based Jupyter Notebook environment called Kaggle Kernels, and using laptop properties processor Core i5, Ram 8 GB, Windows 10.

## 5. Results

### 5.1 Evaluation:

- **Confusion matrix** including: (True Positive TP, True Negative TN, False Positive FP, and False Negative FN).

- **Accuracy**= $\frac{TP+TN}{TP+TN+FP+FN}$

- **Recall**= $\frac{TP}{TP+FN}$

- **Precision**= $\frac{TP}{TP+FP}$

- **Sensitivity**= $\frac{TP}{(TP+FP)}$

- **Specificity**= $\frac{TN}{(TN+FP)}$

- **F1-Score or F-Measure**= $2*\frac{(Recall*Precision)}{(Recall+Precision)}$

### 5.2 System walkthrough:

Figure 6 shows the output of the pre-processing stage. First, the input image is provided. Second, the bounding box of ROI is determined. Third, the output of ROI is zoomed.

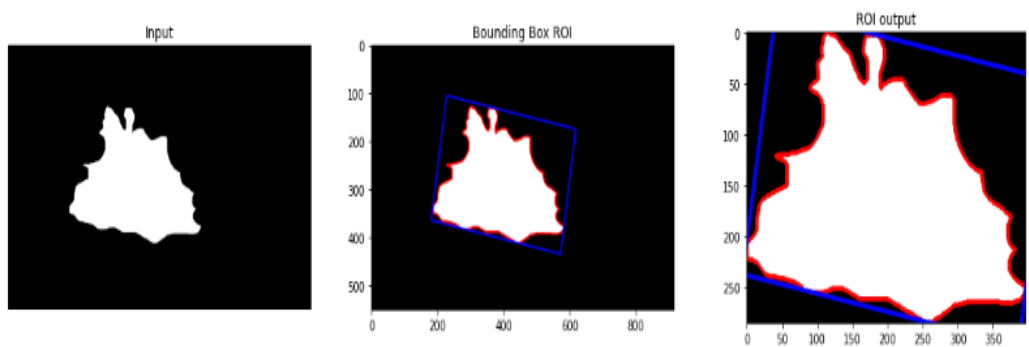

Figure 6: the output of the pre-processing phase

As shown in figure 7, real image, mask of tumor, and mirror of mask, can be easily calculated.

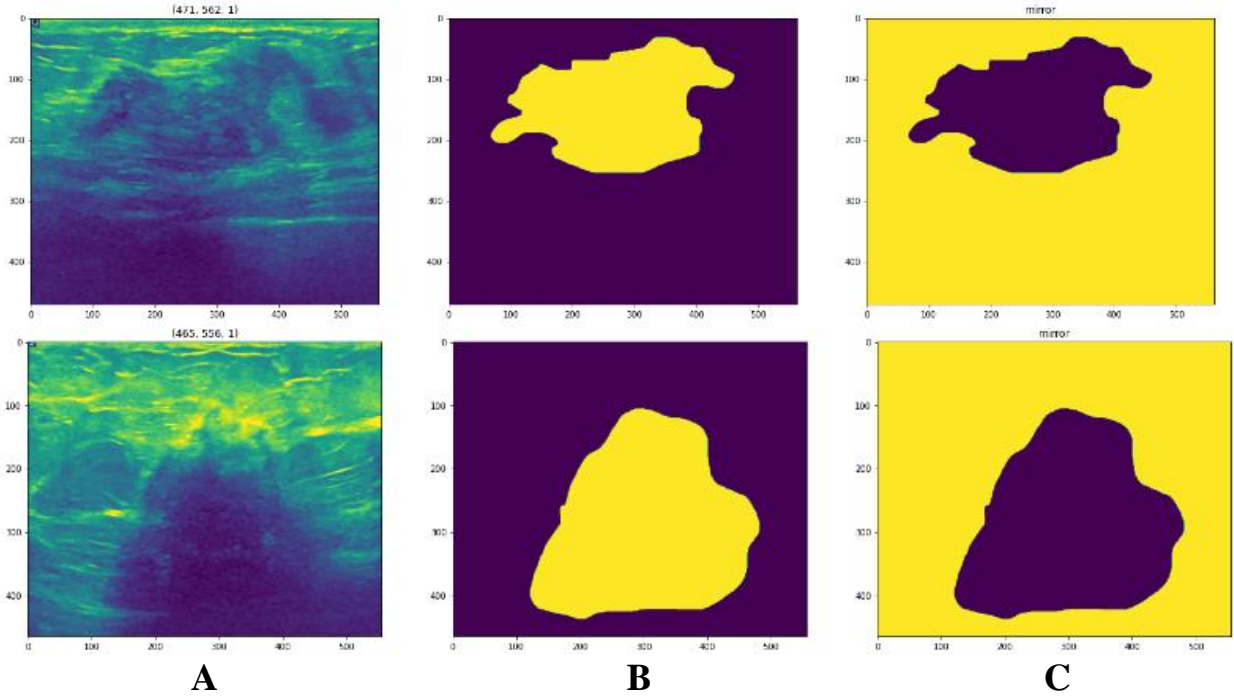

Figure 7: (A) Main image, (B) Mask of image, (C) mirror of mask.

Table 2 shows the training parameters. We have 75 epochs, with batch size is 32. This batch size follows the recommendations of deep learning [34]. The learning rate is $1e^{-2}$.

Table 2: the training parameters

| Epoch | Batch size | learning rate |
|---|---|---|
| 75 | 32 | $1e^{-2}$ |

Table 3 shows the accuracy, loss, and AUC of the training data. Accuracy measures the model's ability to predict correct labels or classes for training data, often used in classification tasks. Loss measures the discrepancy between predicted outputs and actual labels in training data, used to guide the model's learning process. AUC (Area Under the Curve) is a metric used in binary classification tasks to measure and evaluate the model's performance in terms of true positive rate (sensitivity) and false positive rate (1-specificity). A higher AUC indicates improved discrimination ability, with values falling between 0 and 1.

Table 3: the accuracy, loss, and AUC of the training.

| Accuracy | Loss | AUC |
|---|---|---|
| 0.9868 | 0.0338 | 0.9978 |

Figure 8 shows explains the loss of train data. The training loss is calculated by averaging individual losses across the dataset, aiming to minimize it using optimization algorithms like gradient descent, thereby learning patterns that generalize well to unseen data.

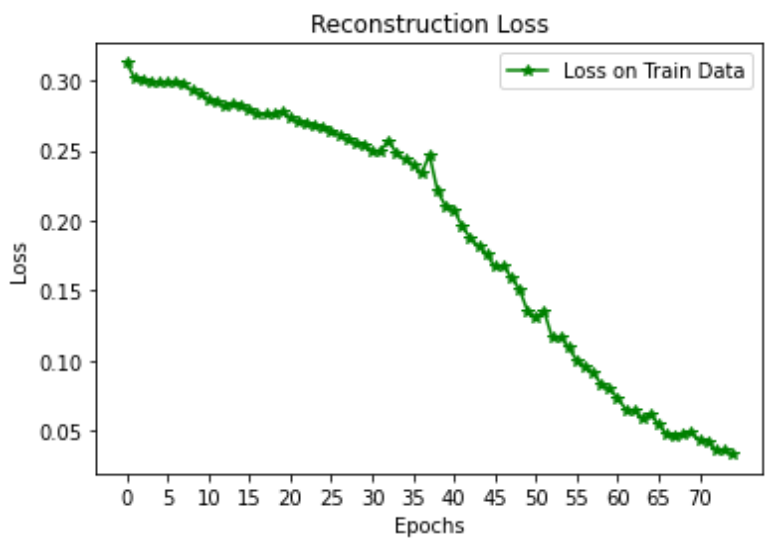

Figure 8: the loss of train data.

Figure 9 shows the main image, real mask, and predicted mask. A mask in computer vision is a binary image that indicates specific regions or objects of interest, used for object segmentation or image annotation. Real masks are manually annotated, representing the true segmentation of objects or regions, while predicted masks are generated by AI models or algorithms, attempting to automatically segment objects or regions in the main image.

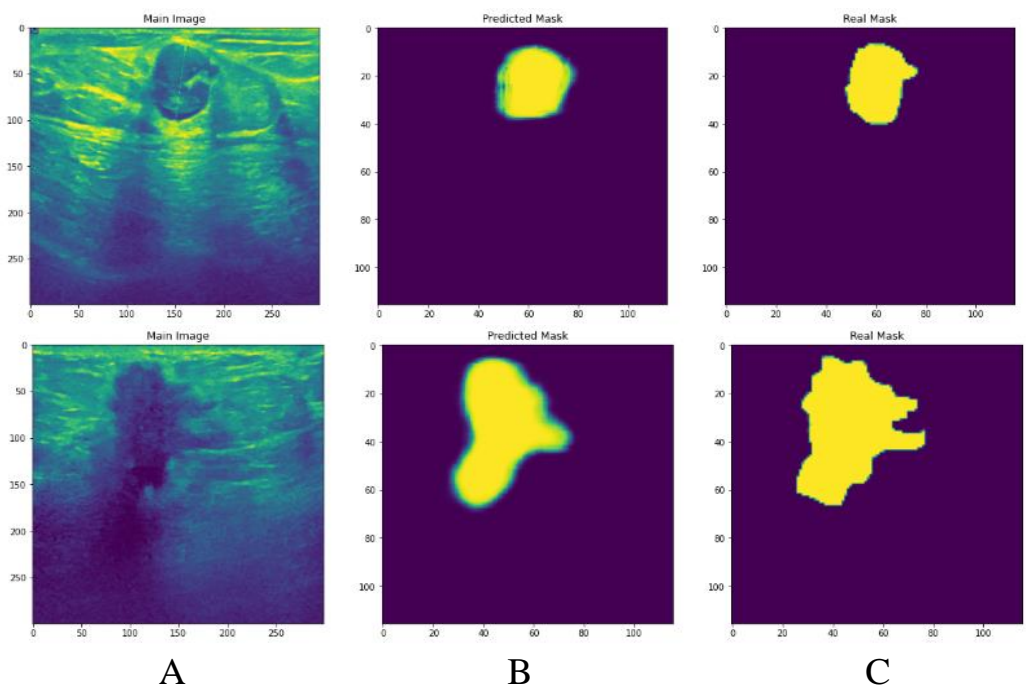

Figure 9: (A) the main image, (B) the predicted mask, (C) the real mask.

Table 4: the layer, parameters, and output shape of VGG classification.

Table 4: layer, parameters, and output shape for VGG.

| Layer (type) | Output Shape | Parameters |
|---|---|---|
| Flatten_1 (Flatten) | (None, 25088) | 0 |
| Batch_normalization_ 1(Batch) | (None, 25088) | 100352 |
| Dense_2 (Dense) | (None, 128) | 3211392 |
| Dense_3 (Dense3) | (None, 3) | 387 |
| **Total parameters:** 3,312,131 **Trainable parameters:** 3,261,955 **Non-trainable parameters:** 50,176 | | |

Table 5 shows the testing parameters. We have 5 epochs, with batch size is 32. This batch size follows the recommendations of deep learning [40]. The learning rate is 0.001.

Table 5: the testing parameters.

| Epochs | Batch size | Learning rate |
|---|---|---|
| 5 | 32 | 0.001 |

Table 6 show the accuracy, validation accuracy, and loss for testing. We have 99.29% accuracy, with validation accuracy 90.48%. The loss is 0.0453.

Table 6: the accuracy, validation accuracy, and loss for testing.

| Accuracy: | Validation accuracy: | loss: |
|---|---|---|
| 0.9929 | 0.9048 | 0.0453 |

Figure 10 shows Classification Model Performance. The loss started from 1.4 and is decreased to reach 0.0453. The accuracy started from beneath 0.8 and reached 0.9929.

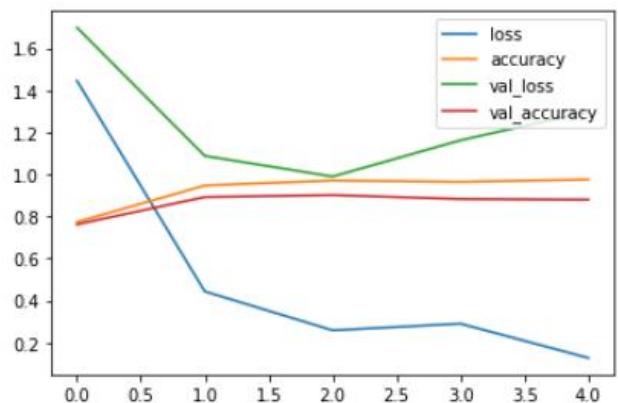

Figure 10: Classification Model Performance

Figure 11 shows ability the proposed system to detect if an image belongs to which class, to be normal, malignant or benign.

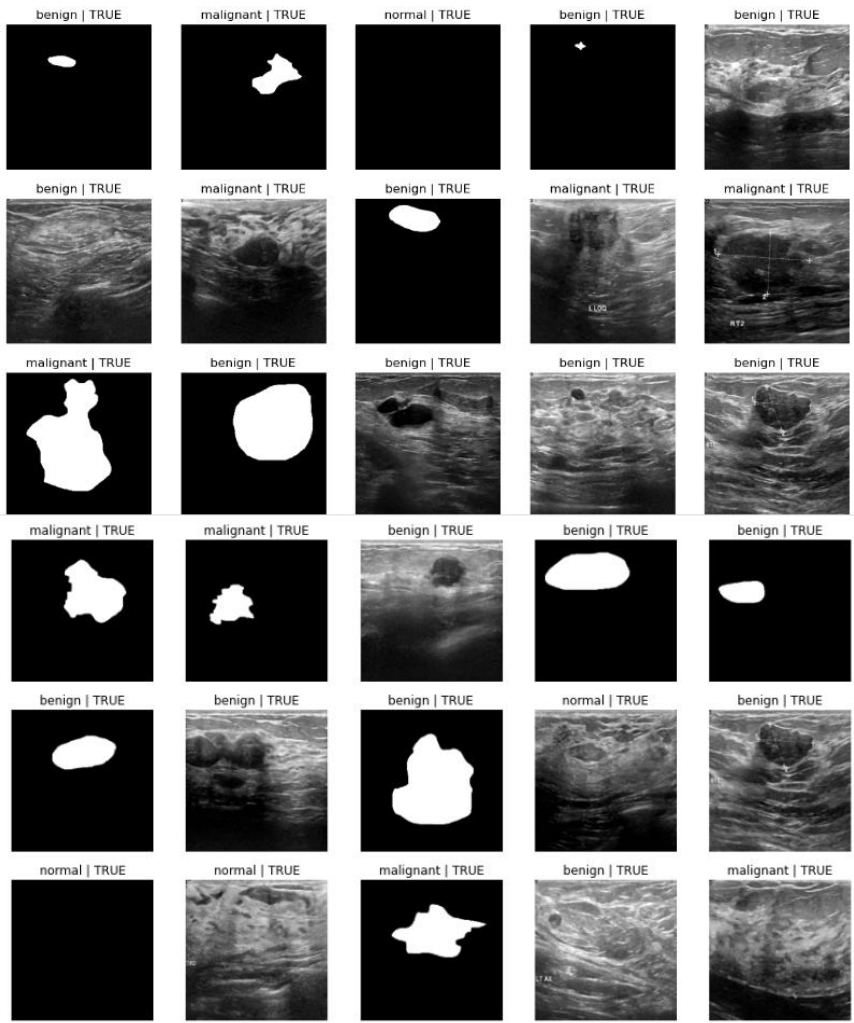

Figure 11: predicting normal, malignant, or benign.

Figure 12 shows only those pictures that are predicted false. Rest of the pictures are predicted true. it's just for clarification of the results.

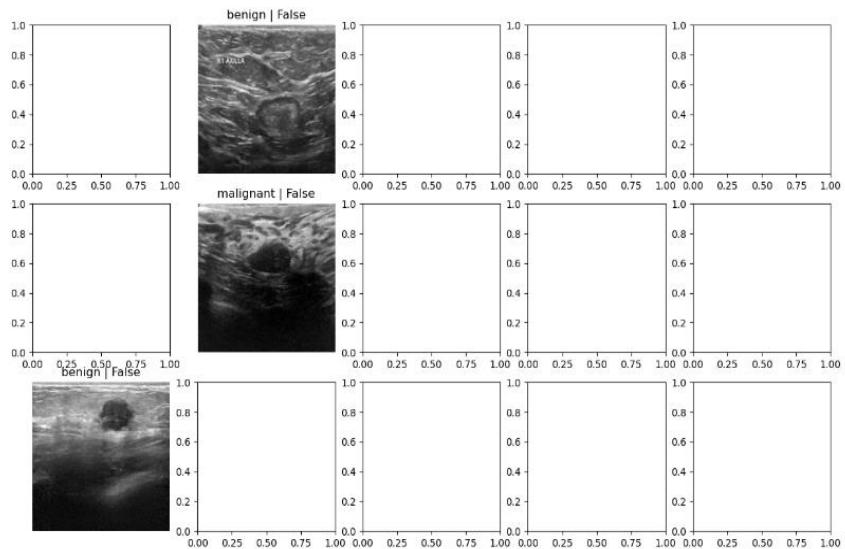

Figure 12: Those only pictures that are predicted false.

Figure 13 shows the confusion matrix. It's a tabular representation that shows the predicted and actual class labels for a set of data, providing a detailed analysis pertaining to the performance of a classification model, especially in scenarios with imbalanced classes. It consists of four key elements: True Positives (TP), True Negatives (TN), False Positives (FP), and False Negatives (FN).

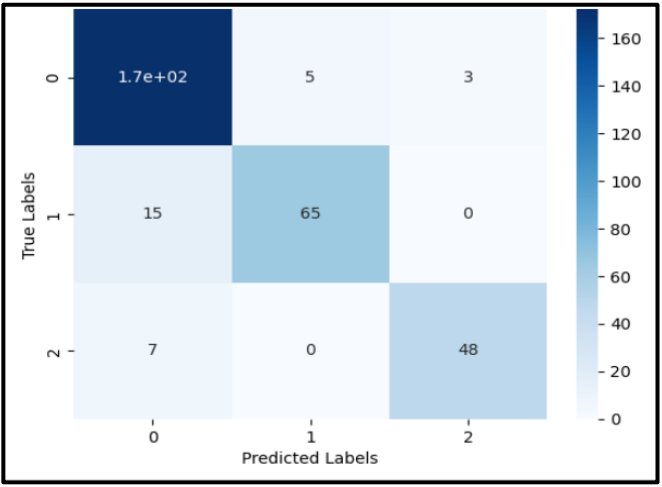

Figure 13: the confusion matrix.

Table 7 lists the measurement of proposed system. The precision is 91.87%. sensitivity is 88.02% while specificity is 93.47%. The recall is 88.02% and the F1-score is 89.73%.

Table 7: the measurement of proposed system.

| Precision | Sensitivity | Specificity | Recall | F1-Score (F-Measure) |
|---|---|---|---|---|
| 0.9187 | 0.8802 | 0.9347 | 0.8802 | 0.8973 |

Table 8 shows the comparison. The proposed model exceeds the previous model in accuracy. It has 99.29% accuracy.

Table 8: comparison proposed system with previous work.

| References | Accuracy |
|---|---|
| [9] | 92.95 |
| [10] | 73.0 |
| [11] | 94.62 |
| [12] | 94.23 |
| [13] | 94.62 |
| [14] | 89.73 |
| [15] | 90.41 |
| [16] | 95.48 |
| [17] | 91.5 |
| [18] | 97 |
| [19] | 97.17 |
| [20] | 95 |
| [21] | 90 |
| [22] | 88.67 |
| [23] | 89.8 |
| **Proposed system** | **99.29** |

## 6. Conclusion and Future directions:

This work proposed a deep learning model that increased the accuracy of ultrasound image-based breast cancer classification. It reaches 99.29% accuracy, comparison with previous work. Used ultrasound image dataset Contains benign, malignant, and normal (no cancer). Included proposed system four phases its: pre-processing, image segmentation, feature extraction, and classification.

One possible future direction may be applying other deep learning models. Another possible future direction is to consider other datasets.

## References

1. Sharma, Ganesh N., et al., "Various types and management of breast cancer: an overview". Journal of advanced pharmaceutical technology & research 1.2 (2010): 109.
2. Kamińska, Marzena, et al., "Breast cancer risk factors". Menopause Review/Przegląd Menopauzalny 14.3 (2015): 196-202.
3. Sharma, Ganesh N., et al., "Various types and management of breast cancer: an overview". Journal of advanced pharmaceutical technology & research 1.2 (2010): 109.
4. Jafari, Seyed Hamed, et al., "Breast cancer diagnosis: Imaging techniques and biochemical markers". Journal of cellular physiology 233.7 (2018): 5200-5213.
5. Waks, Adrienne G., and Eric P. Winer, "Breast cancer treatment: a review". Jama 321.3 (2019): 288-300.
6. Thackeray, Rosemary, et al., "Using Twitter for breast cancer prevention: an analysis of breast cancer awareness month". BMC cancer 13 (2013): 1-9.
7. Mahmood, Tariq, et al., "A brief survey on breast cancer diagnostic with deep learning schemes using multi-image modalities". IEEE Access 8 (2020): 165779-165809.
8. Debelee, Taye Girma, et al., "Survey of deep learning in breast cancer image analysis". Evolving Systems 11 (2020): 143-163.
9. LIAO, Wen-Xuan, et al., "Automatic identification of breast ultrasound image based on supervised block-based region segmentation algorithm and features combination migration deep learning model". IEEE journal of biomedical and health informatics, 2019, 24.4: 984-993.
10. Cao, Zhantao, et al., "Breast tumor classification through learning from noisy labeled ultrasound images". Medical Physics 47.3 (2020): 1048-1057.
11. Moon, Woo Kyung, et al., "Computer-aided diagnosis of breast ultrasound images using ensemble learning from convolutional neural networks". Computer methods and programs in biomedicine 190 (2020): 105361.
12. ZHANG, Erlei, et al., "BIRADS features-oriented semi-supervised deep learning for breast ultrasound computer-aided diagnosis". Physics in Medicine & Biology, 2020, 65.12: 125005.
13. MOON, Woo Kyung, et al., "Computer-aided diagnosis of breast ultrasound images using ensemble learning from convolutional neural networks". Computer methods and programs in biomedicine, 2020, 190: 105361.
14. Ilesanmi, Ademola Enitan et al., "A method for segmentation of tumors in breast ultrasound images using the variant enhanced deep learning". Biocybernetics and Biomedical Engineering 41.2 (2021): 802-818.
15. Pang, Ting, et al., "Semi-supervised GAN-based radiomics model for data augmentation in breast ultrasound mass classification". Computer Methods and Programs in Biomedicine 203 (2021): 106018.
16. Zhuang, Zhemin, et al., "Breast ultrasound tumor image classification using image decomposition and fusion based on adaptive multi-model spatial feature fusion". Computer methods and programs in biomedicine 208 (2021): 106221.
17. Byra, Michal. "Breast mass classification with transfer learning based on scaling of deep representations". Biomedical Signal Processing and Control 69 (2021): 102828.
18. Ayana, Gelan, et al., "Transfer learning in breast cancer diagnoses via ultrasound imaging". Cancers 13.4 (2021): 738.
19. WANG, Ke, et al., "Breast ultrasound image segmentation: A coarse-to-fine fusion convolutional neural network". Medical Physics, 2021, 48.8: 4262-4278.
20. Ayana, Gelan, and Se-Woon Choe, "BUVITNET: Breast ultrasound detection via vision transformers", Diagnostics 12.11 (2022): 2654.
21. Zhang, Y., et al., "BUSIS: A Benchmark for Breast Ultrasound Image Segmentation. Healthcare". 2022, 10, 729.
22. KAPLAN, Ela, et al., "Automated BI-RADS classification of lesions using pyramid triple deep feature generator technique on breast ultrasound images". Medical Engineering & Physics, 2022, 108: 103895.
23. Feng, H., et al., "Identifying Malignant Breast Ultrasound Images Using ViT-Patch". Appl. Sci. 2023, 13, 3489. https://doi.org/10.3390/app13063489.

24. Jayaraman, S., et al., "Digital image processing" Recuperado de https://books.google.com.pe/books (2009).
25. Martı́, R., et al., "Breast skin-line segmentation using contour growing", in [Iberian conference on pattern recognition and image analysis], 564–571, Springer (2007).
26. Raba, D., et al., "Breast segmentation with pectoral muscle suppression on digital mammograms", in [Iberian conference on pattern recognition and image analysis ], 471–478, Springer (2005).
27. Li, H., et al., "Improved breast mass segmentation in mammograms with conditional residual u-net", in [Image Analysis for Moving Organ, Breast, and Thoracic Images], 81–89, Springer (2018).
28. Li, S., et al., "Attention dense-u-net for automatic breast mass segmentation in digital mammogram", IEEE Access 7, 59037–59047 (2019).
29. Salau, A. O., & Jain, S., "Feature extraction: a survey of the types, techniques, applications", international conference on signal processing and communication (ICSC) IEEE, 2019, (pp. 158-164).
30. Olaf Ronneberger, et al., "U-Net: Convolutional Networks for Biomedical Image Segmentation", arXiv:1505.04597v1, 2015.
31. Xiao-Xia Yin et al., "U-Net-Based Medical Image Segmentation", Article ID 4189781, https://doi.org/10.1155/2022/4189781, Vol. 2022.
32. Ikram BEN AHMED et al., "Hybrid UNET Model Segmentation for an Early Breast Cancer Detection using Ultrasound Images", doi.org/10.36227/techrxiv.19704895.v1, 2022.
33. Asma Baccouche et al., "Connected-UNets: a deep learning architecture for breast mass segmentation", npj Breast Cancer 2021 7:151; https://doi.org/10.1038/s41523-021-00358-x.
34. F. Moayedi., et al., "Contourlet-based mammography mass classification", ICIAR Lect. Notes Comput. Sci. 2007.
35. Andrew G. Howard et al., "MobileNets: Efficient Convolutional Neural Networks for Mobile Vision Applications", arXiv:1704.04861v1, 2017.
36. Hartatik Muhammad Khoirul Anam: "Comparison of Convolutional Neural Network Architecture on Detection of Helmet Use by Humans", ELINVO (Electronics, Informatics, and Vocational Education), Mei 2023; vol 8(1):44-54 ISSN 2580-6424 (printed), ISSN 2477-2399 (online) DOI: https://doi.org/10.21831/elinvo.v8i1.52104.
37. J. Hays and A. Efros. "IM2GPS: estimating geographic information from a single image". In Proceedings of the IEEE International Conference on Computer Vision and Pattern Recognition, 2008.
38. WANG, Wei, et al., "A novel image classification approach via dense-MobileNet models". Mobile Information Systems, 2020.

39. Y.-L. HUANG, et al., "Computer-aided diagnosis using morphological features for classifying breast lesions on ultrasound", Ultrasound Obstet Gynecol 2008, 32: 565–572.
40. Qingling Zhang et al., "Development and External Validation of a Simple-To-Use Dynamic Nomogram for Predicting Breast Malignancy Based on Ultrasound Morphometric Features: A Retrospective Multicenter Study", Frontiers in Oncology, 2022, Vol. 12, Article 868164.
41. BENGIO, Yoshua. "Practical recommendations for gradient-based training of deep architectures". In: Neural Networks: Tricks of the Trade: Second Edition. Berlin, Heidelberg: Springer Berlin Heidelberg, 2012. p. 437-478.